\documentclass[aps,twocolumn,longbibliography,groupedaddress]{revtex4-2}%
\usepackage{amsfonts}
\usepackage{amsmath,amssymb,epsfig,color}
\usepackage{float}
\usepackage{amsmath}
\usepackage{amssymb}
\usepackage{graphicx}%
\providecommand{\U}[1]{\protect\rule{.1in}{.1in}}
\AtBeginDocument{    \newwrite\bibnotes
\def\bibnotesext{Notes.bib}
\immediate\openout\bibnotes=\jobname\bibnotesext
\immediate\write\bibnotes{@CONTROL{REVTEX41Control}}
\immediate\write\bibnotes{@CONTROL{    apsrev41Control,author="08",editor="1",pages="1",title="0",year="1"}}
\if@filesw
\immediate\write\@auxout{\string\citation{apsrev41Control}}    \fi
}
\providecommand{\U}[1]{\protect\rule{.1in}{.1in}}
\providecommand{\U}[1]{\protect\rule{.1in}{.1in}}

\newcommand*\nbses[1]{{NbS$_{\rm x}$Se$_{\rm 2 -x}$ }}
\begin{document}
\title{Ising superconductivity in monolayer niobium dichalcogenide alloys}
\author{Darshana Wickramaratne}
\affiliation{Center for Computational Materials Science, U.S. Naval Research Laboratory,
Washington, DC 20375, USA}
\author{I.I. Mazin}
\affiliation{Department of Physics and Astronomy, George Mason University, Fairfax, VA
22030, USA}
\affiliation{Quantum Science and Engineering Center, George Mason University, Fairfax, VA
22030, USA}
\date{\today }

\begin{abstract}
NbSe$_{2}$ and NbS$_{2}$ are isostructural two-dimensional materials that
exhibit contrasting superconducting properties when reduced to the single
monolayer limit. Monolayer NbSe$_{2}$ is an Ising superconductor, while there
have been no reports of superconductivity in monolayer NbS$_{2}$.
\nbses~alloys exhibit an intriguing non-monotonic dependence of the
superconducting transition temperature with sulfur content, which has been
interpreted as a manifestation of fractal superconductivity. However, several
key questions about this result are not known: (1) Does the electronic
structure of the alloy differ from the parent compounds, (2) Are spin
fluctuations which have been shown to be prominent in monolayer NbSe$_{2}$
also present in the alloys? Using first-principles calculations, we show that
the density of states at the Fermi level and the proximity to magnetism in
\nbses~alloys are both reduced compared to the parent compound; the former
would decrease the transition temperature while the latter would increase it.
We also show that Se vacancies, which are likely magnetic pair-breaking
defects, may form in large concentrations in NbSe$_{2}$. Based on our results,
we suggest an alternative explanation of the non-monotonic behavior the
superconducting transition temperature in \nbses~alloys, which does not
require the conjecture of multifractality.

\end{abstract}
\maketitle

\section{Introduction}

Ising superconductivity in two-dimensional materials is a rapidly growing
field of theoretical and experimental research
\cite{xi2016Ising,lu2015evidence,zhou2016Ising,mockli2020Ising,sergio2018tuning,isingprx,costanzo2018tunnelling,saito2016superconductivity}%
. The combination of broken-inversion symmetry and strong spin-orbit coupling
present in single monolayers (MLs) of the two-dimensional transition metal
dichalcogenides leads to Fermi surfaces where the pseudospin of the electrons
is perpendicular to the plane of the monolayer and the pseudospin direction
flips between time-reversal invariant points of the Brillouin zone. This has
been experimentally confirmed by establishing, for example in NbSe$_{2}$, that
the superconducting critical field is significantly higher in-plane versus
out-of-plane, and much larger that the Pauli limit \cite{xi2016Ising}. While
there have been extensive phenomenological descriptions of Ising
superconductivity, there are several intriguing material-specific puzzles.

In NbSe$_{2}$, which is the most widely studied Ising superconductor, the
superconducting transition temperature, $T_{c}$, decreases from $\sim$6 K to
$\sim$ 3 $-$ 4 K, when it is reduced from bulk to a single monolayer
\cite{xi2016Ising}. Similar studies conducted on NbS$_{2}$ provide an
intriguing contrast.  In 2H-NbS$_{2}$ $T_{c}$ is $\sim$ 6 K, while superconductivity
has not been observed in bulk 3R-NbS$_{2}$ \cite{guillamon2008superconducting,witteveen2021polytypism}. 
These two polytypes differ in the stacking of
the individual monolayers, while within each ML Nb atoms are in a
trigonal prismatic coordination with the chalcogen atom, similar to NbSe$_{2}
$. Reducing the thickness of NbS$_{2}$ leads to a strong suppression in
$T_{c}$ \cite{yan2019thickness}. Superconductivity has not been found in
ML NbS$_{2}$.

It was recently reported that when ML NbSe$_{2}$ is alloyed with
sulfur, the $T_{c}$ increases up to the sulfur content of $x$=0.2
\cite{zhao2019disorder} in ML \nbses~alloys. For sulfur content greater
than $\sim$ 0.3, the $T_{c}$ was then found to decrease monotonically
\cite{zhao2019disorder} exhibiting qualitatively similar behavior to the bulk
alloys. This non-monotonic change in $T_{c}$ of the ML alloys, which is
in contrast to the monotonic reduction in $T_{c}$ in bulk \nbses~alloys
\cite{luo2017s,sugawara1993anderson}, was interpreted as disorder-induced
enchancement of $T_{c}$ which possibly arises from the multifractality of the
electronic wave functions \cite{feigel2010fractal,burmistrov2013multifractality}. Implicit in this
assumption is that the role of alloying on electronic and Coulomb interactions
is sufficiently weak so as to not impact $T_{c}$ directly. While this is an enticing
consideration, there are several important questions and experimental puzzles
that need to be addressed first, which we briefly outline.

\begin{figure}[h]
\includegraphics[width=8.5cm]{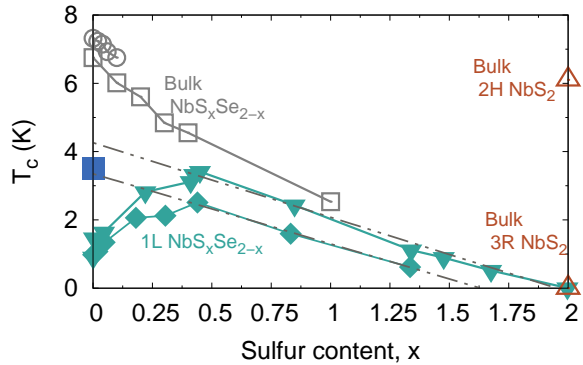}\caption{Experimental reports of the
superconducting transition temperature versus alloy concentration in bulk and
ML \nbses~alloys as a function of sulfur content, $x$. The references
associated with each marker is as follows: $\square$
\cite{sugawara1993anderson}, $\circ$ \cite{luo2017s}, $\triangledown$ and
$\diamond$, \cite{zhao2019disorder}, $\triangle$
\cite{witteveen2021polytypism} and $\blacksquare$ \cite{xi2016Ising}. 
Open symbols correspond to measurements on bulk samples while
filled symbols correspond to measurements on ML samples. The dotted
lines correspond to a linear extrapolation of the ML data for $x>$ 0.3. See the
main text for the discussion on the extrapolation.}%
\label{fig:Tc_x}%
\end{figure}

The measurements where fractal superconductivity was observed report a $T_{c}
$ for ML NbSe$_{2}$ that is $\sim$ 2 K lower than the widely accepted
$T_{c}$ of ML NbSe$_{2}$, $\sim$3 to 4 K
\cite{sergio2018tuning,xi2016Ising}. In fact, the peak $T_{c}$ where fractal
superconductivity is observed is $\sim$ 3 K, which occurs for 0.2 $\leq x\leq$
0.5. We also note the experimental in-plane lattice constant is relatively
unchanged for 0 $\leq x\leq$ 0.2 \cite{SM}. If the $T_{c}$ of
NbSe$_{2}$ in Ref.~\cite{zhao2019disorder} occurred at the more widely
accepted 3 to 4 K, this would not lead to a dome-shaped dependence of $T_{c}$
on sulfur content, as illustrated in Fig.\ref{fig:Tc_x}. Instead, $T_{c}$
would decrease linearly with sulfur content, as has been found when sulfur is
alloyed into bulk NbSe$_{2}$ \cite{luo2017s,sugawara1993anderson}.

Taken together, we arrive at three possible mechanisms that can lead to this
non-monotonic dependence of $T_{c}$ on sulfur content. The first is the role
of fractal superconductivity, which was invoked in
Refs.~\cite{zhao2019disorder,rubio2020visualization}. While this exotic
phenomenon may lead to a non-monotonic change in $T_{c}$
\cite{sacepe2020quantum}, the success of this model requires information on a
plethora of material-dependent parameters that are often not accessible by
experiment alone. A second possible mechanism is the role of the
charge-density wave (CDW), which has been shown to lead to a pseudogapping of
the Fermi surface in ML NbSe$_{2}$
\cite{calandra2009effect,lian2018unveiling}, and thus to a reduction in
$T_{c}$. However, recent studies have suggested the CDW transition temperature
varies little when NbSe$_{2}$ transitions from bulk to a single ML
\cite{ugeda2016characterization,bianco2020weak}, while $T_{c}$ exhibits a
large change \cite{xi2016Ising}. This would imply that the coupling between
the superconducting and CDW order parameters is weak, as has been found in
studies on bulk NbSe$_{2}$ \cite{leroux2015strong,cho2018using}.

A third mechanism is the collective role of point defects
\cite{mockli2020Ising} and spin fluctuations \cite{isingprx}, both of which
have been suggested as a source of pair breaking in ML NbSe$_{2}$.
Experimental studies on ML NbSe$_{2}$ have found the selenium vacancy
concentration can be large (equivalent to a bulk concentration of $\sim
$10$^{21}$ cm$^{-3}$), depending on the growth conditions \cite{wang2017high}.
Selenium vacancies, which are likely magnetic point defects in NbSe$_{2}$
\cite{nbse2_proximity}, can act as a source of pair-breaking and decrease
$T_{c}$. However, during the growth of \nbses~alloys in the presence of
sulfur, which is isovalent to selenium, but more electronegative, it can occupy
the selenium vacancies and lower the concentration of pair-breaking defects.
This is analagous to the finding that oxygen can substitute for sulfur (both
of which are isovalent) in sulfur-deficient ML TaS$_{2}$, and lead to
an increase in $T_{c}$ compared to ML TaS$_{2}$.
\cite{bekaert2020enhanced}.

Alloying will also lead to changes in the electronic structure, which may also
affect the proximity of the material to magnetism or lead to changes in the
density of states (DOS) at the Fermi level, and therefore $T_{c}$. There is
\textit{a priori} no means to determine how all of these properties change
with alloying. Furthermore, if defects are indeed the source of the lower
$T_{c}$ in NbSe$_{2}$, this raises questions on the purported relationship
between the non-monotonic dependence of $T_{c}$ on sulfur content, and fractal
superconductivity \cite{zhao2019disorder}.

In the present work we propose an alternative solution that reconciles these
puzzles. Using first-principles density functional theory calculations
(Sec.~\ref{sec:methods}) we show that this non-monotonic dependence of
$T_{c}$ on sulfur content can emerge from the interplay between defects and
the effect of alloying on the electronic structure and spin-fluctuations. We
show that sulfur is completely miscible in NbSe$_{2}$, across the entire alloy
composition range. For finite concentrations of sulfur in \nbses we find
a reduction of the density of states at the Fermi level \textit{and} a weakening of magnetism,
compared to the parent compounds, NbSe$_{2}$ and NbS$_{2}$. We conjecture a
combination of these effects can lead to a non-monotonic dependence of $T_{c}$
on sulfur content, without having to invoke the phenomenon of multifractality.

\section{Results}

We start by considering the properties of chalcogen vacancies in NbS$_{2}$ and
NbSe$_{2}$ in the dilute limit. The formation energies of a sulfur vacancy,
$V_{\mathrm{S}}$, in NbS$_{2}$ and a selenium vacancy, $V_{\mathrm{Se}}$, in
NbSe$_{2}$ is listed in Table \ref{tab:form}. \begin{table}[h]
\caption{Formation energy of chalcogen vacancies in NbSe$_{2}$ and NbS$_{2}$
under Nb-rich and Nb-poor conditions.}%
\label{tab:form}
\begin{ruledtabular}
\begin{tabular}{ccc}
Defect & Nb-rich (eV) & Nb-poor (eV) \\
\hline
$V_{\rm Se}$ & 0.7 & 1.7 \\
$V_{\rm S}$ & 1.2 & 2 \\
\end{tabular}
\end{ruledtabular}
\end{table}The results show that the formation energy of
$V_{\mathrm{Se}}$ is lower than $V_{\mathrm{S}}$, even under Se-rich
conditions that were used in the growth of the NbSe$_{2}$ samples in Ref. 
\cite{zhao2019disorder}. This suggests
that as-grown ML NbSe$_{2}$ is likely to have a higher concentration of
selenium vacancies compared to sulfur vacancies in NbS$_{2}$. We also
considered the possibility that sulfur may substitute on the Nb site and
calculated the formation energy of this defect, S$_{\mathrm{Nb}}$, in
ML NbSe$_{2}$. In the dilute limit we find the formation energy of
S$_{\mathrm{Nb}}$ to be larger than the formation energy of $V_{\mathrm{Se}}$.
Hence, for the purposes of alloying beyond the dilute limit we only consider
substitution of S on the Se site.

\begin{figure}[h]
\includegraphics[width=8.5cm]{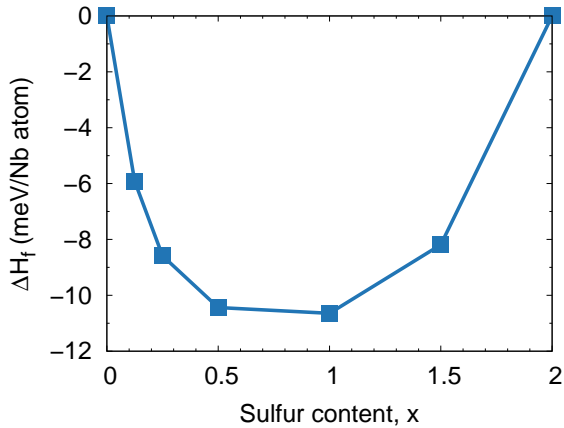}\caption{Formation enthalpy as a function
of sulfur content, $x$, in \nbses.}%
\label{fig:Hx}%
\end{figure}

Next we check the stability of \nbses~alloys with respect to decomposing into
their parent compounds, NbSe$_{2}$ and NbS$_{2}$. Figure~\ref{fig:Hx}
illustrates the lowest enthalpy structure for each composition. We find the
$T$= 0 K formation enthalpy across the entire range of compositions is
negative which suggests ordered \nbses~alloys are stable with respect to decomposition into the parent compounds.

We now turn to the electronic and magnetic properties of the alloys. We first
consider the parent compounds, NbSe$_{2}$ and NbS$_{2}$. In a single ML
the Nb atoms are in a trigonal prismatic coordination with the chalcogen
atoms. In ML NbSe$_{2}$ the trigonal crystal field that acts
on the $4d$ states of Nb$^{4+}$ leads to one band that crosses the Fermi
level, which generates Fermi contours at $\Gamma$, K and K$^{\prime}$
\cite{isingprx}. The combination of broken inversion symmetry in the monolayer
and strong spin-orbit coupling leads to a spin-orbit splitting of the spin
degenerate band along the M-K-$\Gamma$ line of the Brillouin zone. Since the
$4d$ states of Nb$^{4+}$ in NbS$_{2}$ are also in a trigonal prismatic
coordination, albeit with a shorter Nb-S bond length compared to the Nb-Se
bond length, the qualitative features of the band structure between the two
materials are similar \cite{SM}.

ML NbSe$_{2}$ exhibits strong spin fluctuations, which have been
highlighted as a potential source of pair breaking
\cite{isingprx,divilov2021magnetic,das2021renormalized,wan2021unconventional}%
. First-principles calculations have shown that monolayer NbSe$_{2}$ can host
ferromagnetic spin fluctuations with a sizeable Stoner renormalization, and an
antiferromagnetic spin spiral state with \textbf{q} vector (0.2,0,0)
\cite{isingprx,das2021renormalized}. In ML NbSe$_{2}$ we find the spin
spiral state to be 1.7 meV/Nb atom lower in energy compared to the
non-magnetic state. In NbS$_{2}$, we find a spin spiral state 
at a {\bf q}-vector of (0.2,0,0) is also stable \cite{SM}
and is is 1.9 meV/Nb atom lower in energy compared to the non-magnetic ground
state. If spin fluctuations are sizeable in the alloy they can impact pairing
interactions.

\begin{figure*}[th]
\includegraphics[width=18.5cm]{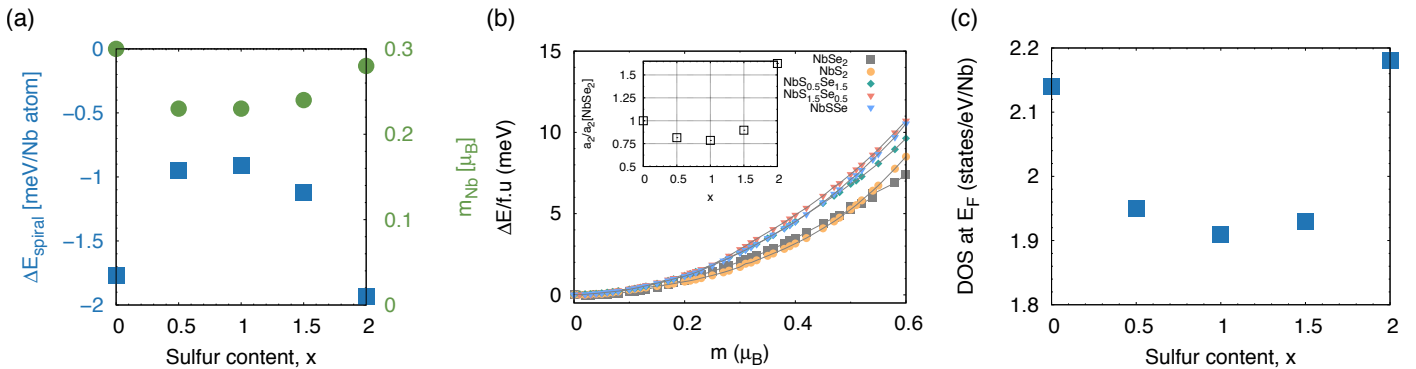}\caption{ (a) Energy
difference between the spin spiral state and the non magnetic state as a
function of sulfur content, $x$, in \nbses~(blue, left vertical axis $-$
$\blacksquare$). Magnetic moment per Nb atom as a function of sulfur content
in the spin spiral calculation with finite $q$ (green, right vertical axis $-$
$\circ$). (b) Collinear fixed-spin moment calculations of NbSe$_{2}$ (grey -
$\blacksquare$), NbS$_{2}$ (orange - $\circ$), NbS$_{0.5}$Se$_{1.5}$ (teal -
$\diamond$), NbSSe (red - $\triangledown$), and NbS$_{1.5}$Se$_{0.5}$ (blue -
$\triangle$) illustrate the change in energy per formula unit with respect to
the non-magnetic state as a function of magnetic moment per Nb atom. The inset
illustrates the coefficient $a_{2}$ (see main text) normalized by the value of
$a_{2}$ in NbSe$_{2}$. (c) Density of states at the Fermi level, $E_{F}$, as a
function of sulfur content, $x$. The magnitude of the DOS for NbSe$_{2}$ and
NbS$_{2}$ correspond to the spin spiral state \cite{SM}. }%
\label{fig:spiral}%
\end{figure*}

To study the effect of alloying on the spin spiral energies we use 
virtual crystal approximation (VCA) calculations (Sec.~\ref{sec:methods}) 
for sulfur contents that correspond to $x$=0.5, 1 and 1.5.
Figure~\ref{fig:spiral}(a)
illustrates the energy difference between the spin spiral state with respect
to the non-magnetic state, $\Delta E_{\mathrm{spiral}}$, in \nbses. When
sulfur is alloyed into NbSe$_{2}$, the spin spiral state is less stable for
intermediate values of sulfur content than for either NbSe$_{2}$ and
NbS$_{2}$.  At $x$=1 we find $\Delta E_{\mathrm{spiral}}$ decreases by a factor of 
2.1 compared to NbS$_2$ where the magnitude of $\Delta E_{\mathrm{spiral}}$ is the largest.
The magnitude of the magnetic moment on the Nb atom is also
suppressed by up to $\simeq25$\%~in the spin-spiral state for the alloys with finite sulfur content compared
to the parent compounds, as illustrated in Fig.~\ref{fig:spiral}%
(a). 

Next we consider whether ferromagnetic spin fluctuations, which are present in NbSe$_2$, are also
impacted due to
alloying by using the VCA and collinear fixed-spin moment calculations for ML
NbSe$_{2}$, NbS$_{2}$, NbSSe, NbS$_{\mathrm{1.5}}$Se$_{\mathrm{0.5}}$, and
NbS$_{\mathrm{0.5}}$Se$_{\mathrm{1.5}}$. We fit these FSM calculations
(illustrated in Fig.~\ref{fig:spiral}(b)) to the following expression,
$E(m)=a_{0}+a_{1}m^{2}+a_{2}m^{4}+a_{3}m^{6}+a_{4}m^{8}$, where $E(m)$ is the
total energy for a given magnetization $m$. The quantity of interest is the
ferromagnetic spin susceptibility, $\chi$, which is defined as $\chi
=a_{2}^{-1}=\left(  \frac{\delta^{2}E}{\delta m^{2}}\right)  ^{-1}$. We find
$\chi$ varies non-monotonically as a function of sulfur content as illustrated
in the inset of Fig.~\ref{fig:spiral}(b), where it is large for NbSe$_{2}$ and
NbS$_{2}$ and suppressed in the case of the alloys. Hence, it is reasonable to
assume that the spin fluctuations for intermediate concentrations are
supressed non-monotonically for all relevant wave vectors. In both cases these
fluctuations are the weakest at roughly equal concentrations of S and Se.

The origin of the reduction in $\chi$ (and, probably, also $\Delta
E_{\mathrm{spiral}},$ given the relatively small spiral vector of (0.2,0,0))
can be understood by examining the density of states (DOS). The DOS at the Fermi level,
$N(\mathrm{E_{F})}$ as a function of sulfur content is illustrated in
Fig.~\ref{fig:spiral}(c). In NbSe$_{2}$ and NbS$_{2}$, $N(E_{F})$ is
suppressed from 2.8 states/eV/Nb atom in the nonmagnetic structure to 2.14
states/eV/Nb atom in NbSe$_2$ and 2.18 states/eV/Nb atom in NbS$_2$ 
in the spin spiral ground state \cite{SM}). 
We also find $N(E_{F})$ is suppressed for the alloys at 
$x$=0.5,1, and 1.5, where in the non-magnetic state $N(E_{F})$ is 
$\sim$2.5 states/eV/Nb atom while in the spin spiral state it reduces to
1.9 states/eV/Nb atom.  Hence, for the parent compounds and the alloys, our calculations
indicate there is a gain in one-electron energy by transitioning to the spin spiral state. 
We also find that in the spin spiral state, the magnitude of $N(E_{F})$
of the alloys decreases by 10\%~compared to $N(E_{F})$ of the parent compounds. 
Such a small change in $N(E_{F})$ as a function of sulfur content is consistent with the fact that
$N(E_{F})$ is comprised almost entirely of Nb $d$-states in NbSe$_2$ and NbS$_2$.

\section{Discussion}
Our calculations lead to two general features that will play an important role
in superconductivity in \nbses~alloys. First, the formation energy of selenium
vacancies in NbSe$_{2}$ is low, which is consistent with large concentrations
of $V_{\mathrm{Se}}$ found in as-grown NbSe$_{2}$. Since they are likely
magnetic \cite{nbse2_proximity}, $V_{\mathrm{Se}}$ can act as a source of
pair-breaking, which would manifest in a reduction in $T_{c}$. This is
consistent with measurements on ML NbSe$_{2}$, where low values of
$T_{c}$ are found in samples where the residual resistivity ratio is low
\cite{wang2017high}. Our calculations show that the substitution of sulfur on
the selenium site in NbSe$_{2}$ is energetically favorable for all sulfur
compositions. Hence, during the growth of \nbses~we expect sulfur to occupy
the sites of missing selenium atoms up to a critical sulfur composition. This
would lower the concentration of pair-breaking $V_{\mathrm{Se}}$ defects. This
immediately explains why the putative multifractal behavior was observed
in NbSe$_{2}$ samples with suppressed $T_c$, compared to the samples in the literature
with lower defect concentrations, which is correlated with higher $T_c$.

Second, we find that both NbSe$_{2}$ and NbS$_{2}$ host strong spin
fluctuations at all wave vectors. However, for finite sulfur compositions in
\nbses~this tendency towards magnetism is weakened, which favors superconductivity.
This reduction in the proximity to magnetism competes with a reduction 
in $N(\mathrm{E_{F})}$ of the alloys compared to NbSe$_{2}$ and
NbS$_{2}$,  which would decrease the electron-phonon coupling constant,
$\lambda_{ep},$ and weaken superconductivity.  We can estimate the sign of the
net effect using the general expression derived in Ref.~\cite{MME}, under a
simplifying assumption that the spin-fluctuations and phonons have comparable
frequencies. Then $d\log T_{c}/d\log\lambda_{ep}\propto\lambda_{ep}%
+2\lambda_{ep}\lambda_{sf},$ and $-d\log T_{c}/d\log\lambda_{sf}\propto
\lambda_{sf}+2\lambda_{sf}\lambda_{ep}.$  The DOS is reduced by
$\approx10\%$ between the end composition and the midpoint.  In contrast, the tendency
to magnetism, as measured by the spin-spiral energy gain, decreases by a factor of 2,
between the end composition and the midpoint.  The latter is expected to be more important,
at least for low concentrations of sulfur.

Taken together, these two features collectively imply that the non-monotonic
dependence of $T_{c}$ on sulfur content\cite{zhao2019disorder} is not
sufficient proof of fractal superconductivity, but likely has a rather prosaic
origin: it is sulfur occupying a high concentration of selenium vacancy sites
(thus decreasing the concentration of pair-breaking defects), which is accompanied by a strong
reduction of the tendency to magnetism as the sulfur concentration increases from
0 to $\sim0.5$. This effect overlaps with the general weaking of the
electron-phonon matrix elements, as evidenced by the smaller coupling constant in
NbS$_{2}$ compared to NbSe$_{2},$ despite their similar $N(\mathrm{E_{F})}$.\cite{heil2017origin,margine,Goran}

We can also show that of the two effects (defects versus reduction in tendency to magnetism)
 that lead to the increase of $T_{c}$
from $x=0$ to $x=0.5$, the former is more important. Indeed, if we linearly
extrapolate the $T_{c}(x)$ data for $x\geq$0.3 to $x=0,$ we get a $T_{c}$ of
NbSe$_{2}$ that ranges from 3.35 K to 4.2 K, as illustrated in
Fig.~\ref{fig:Tc_x}, which is close to the $T_{c}$ reported for ML
NbSe$_{2}$ that has a lower concentration of defects \cite{wang2017high}.

\section{Conclusions}

In conclusion, we have presented a detailed analysis of the properties of
\nbses~alloys using first-principles calculations. Our results, when analyzed
in the context of recent studies that have asserted the presence of fractal
superconductivity in \nbses~alloys
\cite{zhao2019disorder,rubio2020visualization}, suggest multifractality isn't
the only mechanism that can lead to non-monotonic changes in $T_{c}$ in these
alloys. The following key factors emerge from our calculations: (1) the low
formation energy of selenium vacancies that are magnetic pair-breaking point
defects, (2) the stability of \nbses~alloys across the entire composition
range with respect to decomposition into the parent compounds, (3) the
reduction in the density of states at the Fermi level as a function of alloy content
in \nbses~and (4) a reduction in the proximity to magnetism in \nbses~alloys
compared to NbSe$_{2}$ and NbS$_{2}$.

These results suggest that as-grown NbSe$_{2}$ hosts a large concentration of
pair-breaking selenium vacancies that upon alloying are occupied by sulfur
atoms. This leads to an increase in $T_{c}$ up to a critical composition where
the sulfur concentration is equal to concentration of selenium vacancies that
are initially present during the growth. For larger $x,$ $T_{c}$ monotonically
decreases reflecting general weakening of the electron-phonon matrix elements
toward NbS$_{2}$. These two distinct regimes manifest in a non-monotonic
change in $T_{c}$. Given that disorder-induced non-monotonic changes in $T_{c}$
have been observed in other transition metal dichalcogenide alloys due to
isovalent substitution
\cite{peng2018disorder,li2017superconducting,luo2015polytypism}, we expect our
findings to open new avenues for investigation in this broad class of materials.

\section{ Methods}

\label{sec:methods} Our calculations are based on density functional theory
within the projector-augmented wave method \cite{Blochl_PAW} as implemented in
the VASP code \cite{VASP_ref,VASP_ref2} using the generalized gradient
approximation defined by the Perdew-Burke-Ernzerhof (PBE) functional
\cite{perdew1996generalized}. We found it is essential that Nb $5s^{1}%
,4s^{2},4p^{6},4d^{4}$ electrons and Se $4s^{2},4p^{4}$ electrons are treated
as valence. All calculations use a plane-wave energy cutoff of 400~eV. We use
a ($18\times18\times1$) $\Gamma$-centered $k$-point grid for the monolayer
structure when performing structural optimization and calculating the
electronic structure. The cell shape and atomic positions of each
structure was optimized using a force convergence criteria of 5 meV/\AA .

For the defect calculations we use a (10$\times$10$\times$1) supercell of
ML NbSe$_{2}$ and NbSe$_{2}$. To simulate a chalcogen vacancy we remove
a single chalcogen atom (S atom in NbS$_{2}$ and Se atom in NbSe$_{2}$), relax
all of the atomic coordinates and determine the total energy. The formation
energy, of for example, a selenium vacancy, $V_{\mathrm{Se}}$ in NbSe$_{2}$ is
defined as:
\begin{equation}
E^{f}(V_{\mathrm{Se}}) = E_{\mathrm{tot}}(V_{\mathrm{Se}}) - E_{\mathrm{tot}%
}(\mathrm{NbSe_{2}) - \mu_{Se}}%
\end{equation}
where $E^{f}(V_{\mathrm{Se}})$ is the formation energy of the selenium
vacancy, $E_{\mathrm{tot}}(V_{\mathrm{Se}})$ is the total energy of the
NbSe$_{2}$ defect supercell with a selenium vacancy, $E_{\mathrm{tot}%
}(\mathrm{NbSe_{2})}$ is the total energy of the pristine NbSe$_{2}$
supercell, and $\mu_{\mathrm{Se}}$ is the chemical potential of selenium. All
of the defect calculations were performed with a (3$\times$3$\times$1)
$k$-point grid.

For the calculations of the alloy properties we consider two approaches;
the virtual crystal approximation (VCA) and explicit supercell calculations
using either a (4$\times$1$\times$1) and a (4$\times$4$\times$1) supercell that is constructed
from the unit cell of the ML structure.
For each alloy supercell we consider different
arrangements of the S and Se atoms for compositions corresponding to
$x$=0.125, 0.25, 0.5, 1, and 1.5, and relax all of the atomic positions. The
$k$-point grid for structural relaxation of each supercell is scaled with
respect to the ($18\times18\times1$) $\Gamma$-centered $k$-point grid we use
for calculations of the unit cell. 

To determine the thermodynamics of alloy formation we calculated
the formation enthalpy, $\Delta H(x)$, as a function of sulfur content, $x$,
using the (4$\times$4$\times$1) supercell.
$\Delta H(x)$ is defined as:
\begin{equation}
\Delta H(x) = E(x) - xE(\mathrm{NbS_{2}) - (1-x)E(NbSe_{2})}%
\end{equation}
where $E(x)$ is the total energy of the alloy supercell with sulfur content,
$x$, $E(\mathrm{NbS_{2})}$ is the total energy of the NbS$_{2}$ supercell and
$E(\mathrm{NbSe_{2})}$ is the total energy of the NbSe$_{2}$ supercell.

We varied the lattice parameters for each alloy configuration linearly as a
function of sulfur content in accordance with Vegard's law and then relax all of the atomic coordinates. 
For a given sulfur
content, $x$, the in-plane lattice constant, $a$(\nbses~) was varied as
$a$(\nbses~) = $xa_{\mathrm{NbS_{2}}}$ + $(2-x)a_{\mathrm{NbSe_{2}}}$, where
$a_{\mathrm{NbS_{2}}}$ is the in-plane lattice parameter of bulk NbS$_{2}$ and
$a_{\mathrm{NbSe_{2}}}$ is the in-plane lattice parameter of bulk NbSe$_{2}$.
We verified the accuracy of Vegard's law for a subset of alloy structures by
allowing the lattice parameters and atomic positions to relax. In all cases,
the variation of the in-plane lattice parameters was linear \cite{SM}.

To calculate the spin spiral energies we used the generalized
Bloch theorem formalism \cite{sandratskii1991symmetry} as implemented within
VASP. We use a dense ($36\times
36\times1$) $\Gamma$-centered $k$-point grid for the unit cell.
We determine the energy difference between the spin
spiral state with respect to the non-magnetic state, $\Delta
E_{\mathrm{spiral}}$, which is defined as $\Delta E_{\mathrm{spiral}}$ =
$E(\mathbf{q)}$ $- E(\mathbf{q=0)}$ where $E(\mathbf{q)}$ is the total energy
of the unit cell with spin spiral wavevector \textbf{q} and
$E(\mathbf{q=0)}$ is the total energy of the non-magnetic unitcell.

To determine the ferromagnetic spin susceptibility, $\chi$ we used collinear
fixed-spin moment (FSM) calculations (sometimes referred to as the constrained
local moments approach). In our collinear FSM calculations we constrain the
magnitude of the magnetic moment on the Nb atom. Performing these calculations
allows us to determine the change in energy with respect to the non-magnetic
ground state as a function of the total magnetization, $m$. We then fit our results to an expansion of the total energy as a
function of $m$ (see main text to determine $\chi$. The spin susceptibility,
$\chi$, obtained from FSM calculations is sensitive to the choice in energy
convergence threshold, and the number of magnetization values used in the fit
to expansion in the total energy as a function of magnetic moment. We use an
energy convergence threshold of 10$^{-8}$ eV, and up to 50 magnetization
versus energy points between 0 $\mu_{B}$ and 0.6 $\mu_{B}$ for all of the FSM calculations.

The results on the formation enthalpy of the alloys are obtained using a
(4$\times$4$\times$1) supercell with tetrahedron smearing and a ($9\times
9\times1$) $k$-point grid.  The spin spiral energies, fixed spin moment calculations, and
the density of states of the alloys are obtained using VCA calculations
for sulfur contents that correspond to $x$=0.5, 1, and 1.5.
The VCA calculations
use the same $k$-point grid as the unit cell calculations.  The in-plane lattice parameters
for the $x$=0.5, 1, and 1.5 VCA calculations are scaled linearly according to Vegard's law. 
Furthermore, we also interpolate the vertical Nb-chalcogen bond length along the $c$-axis for each
VCA alloy calculation.
%

%\section*{Data Availability}
%The data that supports the findings of this study are available from the corresponding
%author upon reasonable request.

%\section*{Competing Interests}
%The authors declare no competing interests.

\section*{Acknowledgements}
We thank Mikhail Feigel'man and Roxana Margine for helpful discussions. 
D.W was supported by the
Office of Naval Research (ONR) through the Naval Research Laboratory's Basic
Research Program. I.I.M. was supported by ONR through grant N00014-20-1-2345.
Calculations by D.W. were performed at the DoD Major Shared Resource Center at
AFRL. 

%\section*{Author Contributions}
%D.W and I.I.M performed the first-principles calculations, analyzed the results and wrote the paper.

%\bibliography{BIBLIO}

\begin{thebibliography}{43}%
\makeatletter
\providecommand \@ifxundefined [1]{%
 \@ifx{#1\undefined}
}%
\providecommand \@ifnum [1]{%
 \ifnum #1\expandafter \@firstoftwo
 \else \expandafter \@secondoftwo
 \fi
}%
\providecommand \@ifx [1]{%
 \ifx #1\expandafter \@firstoftwo
 \else \expandafter \@secondoftwo
 \fi
}%
\providecommand \natexlab [1]{#1}%
\providecommand \enquote  [1]{``#1''}%
\providecommand \bibnamefont  [1]{#1}%
\providecommand \bibfnamefont [1]{#1}%
\providecommand \citenamefont [1]{#1}%
\providecommand \href@noop [0]{\@secondoftwo}%
\providecommand \href [0]{\begingroup \@sanitize@url \@href}%
\providecommand \@href[1]{\@@startlink{#1}\@@href}%
\providecommand \@@href[1]{\endgroup#1\@@endlink}%
\providecommand \@sanitize@url [0]{\catcode `\\12\catcode `\$12\catcode
  `\&12\catcode `\#12\catcode `\^12\catcode `\_12\catcode `\%12\relax}%
\providecommand \@@startlink[1]{}%
\providecommand \@@endlink[0]{}%
\providecommand \url  [0]{\begingroup\@sanitize@url \@url }%
\providecommand \@url [1]{\endgroup\@href {#1}{\urlprefix }}%
\providecommand \urlprefix  [0]{URL }%
\providecommand \Eprint [0]{\href }%
\providecommand \doibase [0]{https://doi.org/}%
\providecommand \selectlanguage [0]{\@gobble}%
\providecommand \bibinfo  [0]{\@secondoftwo}%
\providecommand \bibfield  [0]{\@secondoftwo}%
\providecommand \translation [1]{[#1]}%
\providecommand \BibitemOpen [0]{}%
\providecommand \bibitemStop [0]{}%
\providecommand \bibitemNoStop [0]{.\EOS\space}%
\providecommand \EOS [0]{\spacefactor3000\relax}%
\providecommand \BibitemShut  [1]{\csname bibitem#1\endcsname}%
\let\auto@bib@innerbib\@empty
%</preamble>
\bibitem [{\citenamefont {Xi}\ \emph {et~al.}(2016)\citenamefont {Xi},
  \citenamefont {Wang}, \citenamefont {Zhao}, \citenamefont {Park},
  \citenamefont {Law}, \citenamefont {Berger}, \citenamefont {Forr{\'o}},
  \citenamefont {Shan},\ and\ \citenamefont {Mak}}]{xi2016Ising}%
  \BibitemOpen
  \bibfield  {author} {\bibinfo {author} {\bibfnamefont {X.}~\bibnamefont
  {Xi}}, \bibinfo {author} {\bibfnamefont {Z.}~\bibnamefont {Wang}}, \bibinfo
  {author} {\bibfnamefont {W.}~\bibnamefont {Zhao}}, \bibinfo {author}
  {\bibfnamefont {J.-H.}\ \bibnamefont {Park}}, \bibinfo {author}
  {\bibfnamefont {K.~T.}\ \bibnamefont {Law}}, \bibinfo {author} {\bibfnamefont
  {H.}~\bibnamefont {Berger}}, \bibinfo {author} {\bibfnamefont
  {L.}~\bibnamefont {Forr{\'o}}}, \bibinfo {author} {\bibfnamefont
  {J.}~\bibnamefont {Shan}},\ and\ \bibinfo {author} {\bibfnamefont {K.~F.}\
  \bibnamefont {Mak}},\ }\href@noop {} {\bibfield  {journal} {\bibinfo
  {journal} {Nat. Phys.}\ }\textbf {\bibinfo {volume} {12}},\ \bibinfo {pages}
  {139} (\bibinfo {year} {2016})}\BibitemShut {NoStop}%
\bibitem [{\citenamefont {Lu}\ \emph {et~al.}(2015)\citenamefont {Lu},
  \citenamefont {Zheliuk}, \citenamefont {Leermakers}, \citenamefont {Yuan},
  \citenamefont {Zeitler}, \citenamefont {Law},\ and\ \citenamefont
  {Ye}}]{lu2015evidence}%
  \BibitemOpen
  \bibfield  {author} {\bibinfo {author} {\bibfnamefont {J.}~\bibnamefont
  {Lu}}, \bibinfo {author} {\bibfnamefont {O.}~\bibnamefont {Zheliuk}},
  \bibinfo {author} {\bibfnamefont {I.}~\bibnamefont {Leermakers}}, \bibinfo
  {author} {\bibfnamefont {N.~F.}\ \bibnamefont {Yuan}}, \bibinfo {author}
  {\bibfnamefont {U.}~\bibnamefont {Zeitler}}, \bibinfo {author} {\bibfnamefont
  {K.~T.}\ \bibnamefont {Law}},\ and\ \bibinfo {author} {\bibfnamefont
  {J.}~\bibnamefont {Ye}},\ }\href@noop {} {\bibfield  {journal} {\bibinfo
  {journal} {Science}\ }\textbf {\bibinfo {volume} {350}},\ \bibinfo {pages}
  {1353} (\bibinfo {year} {2015})}\BibitemShut {NoStop}%
\bibitem [{\citenamefont {Zhou}\ \emph {et~al.}(2016)\citenamefont {Zhou},
  \citenamefont {Yuan}, \citenamefont {Jiang},\ and\ \citenamefont
  {Law}}]{zhou2016Ising}%
  \BibitemOpen
  \bibfield  {author} {\bibinfo {author} {\bibfnamefont {B.~T.}\ \bibnamefont
  {Zhou}}, \bibinfo {author} {\bibfnamefont {N.~F.}\ \bibnamefont {Yuan}},
  \bibinfo {author} {\bibfnamefont {H.-L.}\ \bibnamefont {Jiang}},\ and\
  \bibinfo {author} {\bibfnamefont {K.~T.}\ \bibnamefont {Law}},\ }\href@noop
  {} {\bibfield  {journal} {\bibinfo  {journal} {Phys. Rev. B}\ }\textbf
  {\bibinfo {volume} {93}},\ \bibinfo {pages} {180501} (\bibinfo {year}
  {2016})}\BibitemShut {NoStop}%
\bibitem [{\citenamefont {M{\"o}ckli}\ and\ \citenamefont
  {Khodas}(2020)}]{mockli2020Ising}%
  \BibitemOpen
  \bibfield  {author} {\bibinfo {author} {\bibfnamefont {D.}~\bibnamefont
  {M{\"o}ckli}}\ and\ \bibinfo {author} {\bibfnamefont {M.}~\bibnamefont
  {Khodas}},\ }\href@noop {} {\bibfield  {journal} {\bibinfo  {journal} {Phys.
  Rev. B}\ }\textbf {\bibinfo {volume} {101}},\ \bibinfo {pages} {014510}
  (\bibinfo {year} {2020})}\BibitemShut {NoStop}%
\bibitem [{\citenamefont {Sergio}\ \emph {et~al.}(2018)\citenamefont {Sergio},
  \citenamefont {Sinko}, \citenamefont {Gopalan}, \citenamefont {Sivadas},
  \citenamefont {Seyler}, \citenamefont {Watanabe}, \citenamefont {Taniguchi},
  \citenamefont {Tsen}, \citenamefont {Xu}, \citenamefont {Xiao},\ and\
  \citenamefont {Hunt}}]{sergio2018tuning}%
  \BibitemOpen
  \bibfield  {author} {\bibinfo {author} {\bibfnamefont {C.}~\bibnamefont
  {Sergio}}, \bibinfo {author} {\bibfnamefont {M.~R.}\ \bibnamefont {Sinko}},
  \bibinfo {author} {\bibfnamefont {D.~P.}\ \bibnamefont {Gopalan}}, \bibinfo
  {author} {\bibfnamefont {N.}~\bibnamefont {Sivadas}}, \bibinfo {author}
  {\bibfnamefont {K.~L.}\ \bibnamefont {Seyler}}, \bibinfo {author}
  {\bibfnamefont {K.}~\bibnamefont {Watanabe}}, \bibinfo {author}
  {\bibfnamefont {T.}~\bibnamefont {Taniguchi}}, \bibinfo {author}
  {\bibfnamefont {A.~W.}\ \bibnamefont {Tsen}}, \bibinfo {author}
  {\bibfnamefont {X.}~\bibnamefont {Xu}}, \bibinfo {author} {\bibfnamefont
  {D.}~\bibnamefont {Xiao}},\ and\ \bibinfo {author} {\bibfnamefont
  {B.}~\bibnamefont {Hunt}},\ }\href@noop {} {\bibfield  {journal} {\bibinfo
  {journal} {Nat. Comm.}\ }\textbf {\bibinfo {volume} {9}},\ \bibinfo {pages}
  {1427} (\bibinfo {year} {2018})}\BibitemShut {NoStop}%
\bibitem [{\citenamefont {Wickramaratne}\ \emph {et~al.}(2020)\citenamefont
  {Wickramaratne}, \citenamefont {Khmelevskyi}, \citenamefont {Agterberg},\
  and\ \citenamefont {Mazin}}]{isingprx}%
  \BibitemOpen
  \bibfield  {author} {\bibinfo {author} {\bibfnamefont {D.}~\bibnamefont
  {Wickramaratne}}, \bibinfo {author} {\bibfnamefont {S.}~\bibnamefont
  {Khmelevskyi}}, \bibinfo {author} {\bibfnamefont {D.~F.}\ \bibnamefont
  {Agterberg}},\ and\ \bibinfo {author} {\bibfnamefont {I.}~\bibnamefont
  {Mazin}},\ }\href@noop {} {\bibfield  {journal} {\bibinfo  {journal}
  {Phys. Rev. X}\ }\textbf {\bibinfo {volume} {10}},\ \bibinfo {pages}
  {041003} (\bibinfo {year} {2020})}\BibitemShut {NoStop}%
\bibitem [{\citenamefont {Costanzo}\ \emph {et~al.}(2018)\citenamefont
  {Costanzo}, \citenamefont {Zhang}, \citenamefont {Reddy}, \citenamefont
  {Berger},\ and\ \citenamefont {{M}orpurgo}}]{costanzo2018tunnelling}%
  \BibitemOpen
  \bibfield  {author} {\bibinfo {author} {\bibfnamefont {D.}~\bibnamefont
  {Costanzo}}, \bibinfo {author} {\bibfnamefont {H.}~\bibnamefont {Zhang}},
  \bibinfo {author} {\bibfnamefont {B.~A.}\ \bibnamefont {Reddy}}, \bibinfo
  {author} {\bibfnamefont {H.}~\bibnamefont {Berger}},\ and\ \bibinfo {author}
  {\bibfnamefont {A.~F.}\ \bibnamefont {{M}orpurgo}},\ }\href@noop {}
  {\bibfield  {journal} {\bibinfo  {journal} {Nat. Nano}\ }\textbf {\bibinfo
  {volume} {13}},\ \bibinfo {pages} {483} (\bibinfo {year} {2018})}\BibitemShut
  {NoStop}%
\bibitem [{\citenamefont {Saito}\ \emph {et~al.}(2016)\citenamefont {Saito},
  \citenamefont {Nakamura}, \citenamefont {Bahramy}, \citenamefont {Kohama},
  \citenamefont {Ye}, \citenamefont {Kasahara}, \citenamefont {Nakagawa},
  \citenamefont {Onga}, \citenamefont {Tokunaga}, \citenamefont {Nojima} \emph
  {et~al.}}]{saito2016superconductivity}%
  \BibitemOpen
  \bibfield  {author} {\bibinfo {author} {\bibfnamefont {Y.}~\bibnamefont
  {Saito}}, \bibinfo {author} {\bibfnamefont {Y.}~\bibnamefont {Nakamura}},
  \bibinfo {author} {\bibfnamefont {M.~S.}\ \bibnamefont {Bahramy}}, \bibinfo
  {author} {\bibfnamefont {Y.}~\bibnamefont {Kohama}}, \bibinfo {author}
  {\bibfnamefont {J.}~\bibnamefont {Ye}}, \bibinfo {author} {\bibfnamefont
  {Y.}~\bibnamefont {Kasahara}}, \bibinfo {author} {\bibfnamefont
  {Y.}~\bibnamefont {Nakagawa}}, \bibinfo {author} {\bibfnamefont
  {M.}~\bibnamefont {Onga}}, \bibinfo {author} {\bibfnamefont {M.}~\bibnamefont
  {Tokunaga}}, \bibinfo {author} {\bibfnamefont {T.}~\bibnamefont {Nojima}},
  \emph {et~al.},\ }\href@noop {} {\bibfield  {journal} {\bibinfo  {journal}
  {Nat. Phys.}\ }\textbf {\bibinfo {volume} {12}},\ \bibinfo {pages} {144}
  (\bibinfo {year} {2016})}\BibitemShut {NoStop}%
\bibitem [{\citenamefont {Guillam{\'o}n}\ \emph {et~al.}(2008)\citenamefont
  {Guillam{\'o}n}, \citenamefont {Suderow}, \citenamefont {Vieira},
  \citenamefont {Cario}, \citenamefont {Diener},\ and\ \citenamefont
  {Rodiere}}]{guillamon2008superconducting}%
  \BibitemOpen
  \bibfield  {author} {\bibinfo {author} {\bibfnamefont {I.}~\bibnamefont
  {Guillam{\'o}n}}, \bibinfo {author} {\bibfnamefont {H.}~\bibnamefont
  {Suderow}}, \bibinfo {author} {\bibfnamefont {S.}~\bibnamefont {Vieira}},
  \bibinfo {author} {\bibfnamefont {L.}~\bibnamefont {Cario}}, \bibinfo
  {author} {\bibfnamefont {P.}~\bibnamefont {Diener}},\ and\ \bibinfo {author}
  {\bibfnamefont {P.}~\bibnamefont {Rodiere}},\ }\href@noop {} {\bibfield
  {journal} {\bibinfo  {journal} {Phys. Rev. Lett.}\ }\textbf {\bibinfo
  {volume} {101}},\ \bibinfo {pages} {166407} (\bibinfo {year}
  {2008})}\BibitemShut {NoStop}%
\bibitem [{\citenamefont {Witteveen}\ \emph {et~al.}(2021)\citenamefont
  {Witteveen}, \citenamefont {G{\'o}rnicka}, \citenamefont {Chang},
  \citenamefont {M{\aa}nsson}, \citenamefont {Klimczuk},\ and\ \citenamefont
  {von Rohr}}]{witteveen2021polytypism}%
  \BibitemOpen
  \bibfield  {author} {\bibinfo {author} {\bibfnamefont {C.}~\bibnamefont
  {Witteveen}}, \bibinfo {author} {\bibfnamefont {K.}~\bibnamefont
  {G{\'o}rnicka}}, \bibinfo {author} {\bibfnamefont {J.}~\bibnamefont {Chang}},
  \bibinfo {author} {\bibfnamefont {M.}~\bibnamefont {M{\aa}nsson}}, \bibinfo
  {author} {\bibfnamefont {T.}~\bibnamefont {Klimczuk}},\ and\ \bibinfo
  {author} {\bibfnamefont {F.~O.}\ \bibnamefont {von Rohr}},\ }\href@noop {}
  {\bibfield  {journal} {\bibinfo  {journal} {Dalton Transactions}\ }\textbf
  {\bibinfo {volume} {50}},\ \bibinfo {pages} {3216} (\bibinfo {year}
  {2021})}\BibitemShut {NoStop}%
\bibitem [{\citenamefont {Yan}\ \emph {et~al.}(2019)\citenamefont {Yan},
  \citenamefont {Khalsa}, \citenamefont {Schaefer}, \citenamefont {Jarjour},
  \citenamefont {Rouvimov}, \citenamefont {Nowack}, \citenamefont {Xing},\ and\
  \citenamefont {Jena}}]{yan2019thickness}%
  \BibitemOpen
  \bibfield  {author} {\bibinfo {author} {\bibfnamefont {R.}~\bibnamefont
  {Yan}}, \bibinfo {author} {\bibfnamefont {G.}~\bibnamefont {Khalsa}},
  \bibinfo {author} {\bibfnamefont {B.~T.}\ \bibnamefont {Schaefer}}, \bibinfo
  {author} {\bibfnamefont {A.}~\bibnamefont {Jarjour}}, \bibinfo {author}
  {\bibfnamefont {S.}~\bibnamefont {Rouvimov}}, \bibinfo {author}
  {\bibfnamefont {K.~C.}\ \bibnamefont {Nowack}}, \bibinfo {author}
  {\bibfnamefont {H.~G.}\ \bibnamefont {Xing}},\ and\ \bibinfo {author}
  {\bibfnamefont {D.}~\bibnamefont {Jena}},\ }\href@noop {} {\bibfield
  {journal} {\bibinfo  {journal} {Applied Physics Express}\ }\textbf {\bibinfo
  {volume} {12}},\ \bibinfo {pages} {023008} (\bibinfo {year}
  {2019})}\BibitemShut {NoStop}%
\bibitem [{\citenamefont {Zhao}\ \emph {et~al.}(2019)\citenamefont {Zhao},
  \citenamefont {Lin}, \citenamefont {Xiao}, \citenamefont {Huang},
  \citenamefont {Yao}, \citenamefont {Yan}, \citenamefont {Xing}, \citenamefont
  {Zhang}, \citenamefont {Li}, \citenamefont {Hoshino} \emph
  {et~al.}}]{zhao2019disorder}%
  \BibitemOpen
  \bibfield  {author} {\bibinfo {author} {\bibfnamefont {K.}~\bibnamefont
  {Zhao}}, \bibinfo {author} {\bibfnamefont {H.}~\bibnamefont {Lin}}, \bibinfo
  {author} {\bibfnamefont {X.}~\bibnamefont {Xiao}}, \bibinfo {author}
  {\bibfnamefont {W.}~\bibnamefont {Huang}}, \bibinfo {author} {\bibfnamefont
  {W.}~\bibnamefont {Yao}}, \bibinfo {author} {\bibfnamefont {M.}~\bibnamefont
  {Yan}}, \bibinfo {author} {\bibfnamefont {Y.}~\bibnamefont {Xing}}, \bibinfo
  {author} {\bibfnamefont {Q.}~\bibnamefont {Zhang}}, \bibinfo {author}
  {\bibfnamefont {Z.-X.}\ \bibnamefont {Li}}, \bibinfo {author} {\bibfnamefont
  {S.}~\bibnamefont {Hoshino}}, \emph {et~al.},\ }\href@noop {} {\bibfield
  {journal} {\bibinfo  {journal} {Nat. Phys.}\ }\textbf {\bibinfo {volume}
  {15}},\ \bibinfo {pages} {904} (\bibinfo {year} {2019})}\BibitemShut
  {NoStop}%
\bibitem [{\citenamefont {Luo}\ \emph {et~al.}(2017)\citenamefont {Luo},
  \citenamefont {Strychalska-Nowak}, \citenamefont {Li}, \citenamefont {Tao},
  \citenamefont {Klimczuk},\ and\ \citenamefont {Cava}}]{luo2017s}%
  \BibitemOpen
  \bibfield  {author} {\bibinfo {author} {\bibfnamefont {H.}~\bibnamefont
  {Luo}}, \bibinfo {author} {\bibfnamefont {J.}~\bibnamefont
  {Strychalska-Nowak}}, \bibinfo {author} {\bibfnamefont {J.}~\bibnamefont
  {Li}}, \bibinfo {author} {\bibfnamefont {J.}~\bibnamefont {Tao}}, \bibinfo
  {author} {\bibfnamefont {T.}~\bibnamefont {Klimczuk}},\ and\ \bibinfo
  {author} {\bibfnamefont {R.~J.}\ \bibnamefont {Cava}},\ }\href@noop {}
  {\bibfield  {journal} {\bibinfo  {journal} {Chemistry of Materials}\ }\textbf
  {\bibinfo {volume} {29}},\ \bibinfo {pages} {3704} (\bibinfo {year}
  {2017})}\BibitemShut {NoStop}%
\bibitem [{\citenamefont {Sugawara}\ \emph {et~al.}(1993)\citenamefont
  {Sugawara}, \citenamefont {Yokota}, \citenamefont {Takemoto}, \citenamefont
  {Tanokura},\ and\ \citenamefont {Sekine}}]{sugawara1993anderson}%
  \BibitemOpen
  \bibfield  {author} {\bibinfo {author} {\bibfnamefont {K.}~\bibnamefont
  {Sugawara}}, \bibinfo {author} {\bibfnamefont {K.}~\bibnamefont {Yokota}},
  \bibinfo {author} {\bibfnamefont {J.}~\bibnamefont {Takemoto}}, \bibinfo
  {author} {\bibfnamefont {Y.}~\bibnamefont {Tanokura}},\ and\ \bibinfo
  {author} {\bibfnamefont {T.}~\bibnamefont {Sekine}},\ }\href@noop {}
  {\bibfield  {journal} {\bibinfo  {journal} {Journal of low temperature
  physics}\ }\textbf {\bibinfo {volume} {91}},\ \bibinfo {pages} {39} (\bibinfo
  {year} {1993})}\BibitemShut {NoStop}%
\bibitem [{\citenamefont {Feigel'man}\ \emph {et~al.}(2010)\citenamefont
  {Feigel'man}, \citenamefont {Ioffe}, \citenamefont {Kravtsov},\ and\
  \citenamefont {Cuevas}}]{feigel2010fractal}%
  \BibitemOpen
  \bibfield  {author} {\bibinfo {author} {\bibfnamefont {M.}~\bibnamefont
  {Feigel'man}}, \bibinfo {author} {\bibfnamefont {L.}~\bibnamefont {Ioffe}},
  \bibinfo {author} {\bibfnamefont {V.}~\bibnamefont {Kravtsov}},\ and\
  \bibinfo {author} {\bibfnamefont {E.}~\bibnamefont {Cuevas}},\ }\href@noop {}
  {\bibfield  {journal} {\bibinfo  {journal} {Annals of Physics}\ }\textbf
  {\bibinfo {volume} {325}},\ \bibinfo {pages} {1390} (\bibinfo {year}
  {2010})}\BibitemShut {NoStop}%
\bibitem [{\citenamefont {Burmistrov}\ \emph {et~al.}(2013)\citenamefont
  {Burmistrov}, \citenamefont {Gornyi},\ and\ \citenamefont
  {Mirlin}}]{burmistrov2013multifractality}%
  \BibitemOpen
  \bibfield  {author} {\bibinfo {author} {\bibfnamefont {I.}~\bibnamefont
  {Burmistrov}}, \bibinfo {author} {\bibfnamefont {I.}~\bibnamefont {Gornyi}},\
  and\ \bibinfo {author} {\bibfnamefont {A.}~\bibnamefont {Mirlin}},\
  }\href@noop {} {\bibfield  {journal} {\bibinfo  {journal} {Phys. Rev. Lett.}\
  }\textbf {\bibinfo {volume} {111}},\ \bibinfo {pages} {066601} (\bibinfo
  {year} {2013})}\BibitemShut {NoStop}%
\bibitem [{SM()}]{SM}%
  \BibitemOpen
  \href@noop {} {}\bibinfo {note} {See Supplemental Material at [url] for
  additional details.}\BibitemShut {Stop}%
\bibitem [{\citenamefont {Rubio-Verdu}\ \emph {et~al.}(2020)\citenamefont
  {Rubio-Verdu}, \citenamefont {Garcia-Garcia}, \citenamefont {Ryu},
  \citenamefont {Choi}, \citenamefont {Zaldivar}, \citenamefont {Tang},
  \citenamefont {Fan}, \citenamefont {Shen}, \citenamefont {{M}o},
  \citenamefont {Pascual},\ and\ \citenamefont
  {Ugeda}}]{rubio2020visualization}%
  \BibitemOpen
  \bibfield  {author} {\bibinfo {author} {\bibfnamefont {C.}~\bibnamefont
  {Rubio-Verdu}}, \bibinfo {author} {\bibfnamefont {A.~M.}\ \bibnamefont
  {Garcia-Garcia}}, \bibinfo {author} {\bibfnamefont {H.}~\bibnamefont {Ryu}},
  \bibinfo {author} {\bibfnamefont {D.-J.}\ \bibnamefont {Choi}}, \bibinfo
  {author} {\bibfnamefont {J.}~\bibnamefont {Zaldivar}}, \bibinfo {author}
  {\bibfnamefont {S.}~\bibnamefont {Tang}}, \bibinfo {author} {\bibfnamefont
  {B.}~\bibnamefont {Fan}}, \bibinfo {author} {\bibfnamefont {Z.-X.}\
  \bibnamefont {Shen}}, \bibinfo {author} {\bibfnamefont {S.-K.}\ \bibnamefont
  {{M}o}}, \bibinfo {author} {\bibfnamefont {J.~I.}\ \bibnamefont {Pascual}},\
  and\ \bibinfo {author} {\bibfnamefont {M.~M.}\ \bibnamefont {Ugeda}},\ }\href
  {https://doi.org/10.1021/acs.nanolett.0c01288} {\bibfield  {journal}
  {\bibinfo  {journal} {Nano Letters}\ }\textbf {\bibinfo {volume} {20}},\
  \bibinfo {pages} {5111} (\bibinfo {year} {2020})},\ \bibinfo {note} {pMID:
  32463696},\ \Eprint
  {https://arxiv.org/abs/https://doi.org/10.1021/acs.nanolett.0c01288}
  {https://doi.org/10.1021/acs.nanolett.0c01288} \BibitemShut {NoStop}%
\bibitem [{\citenamefont {Sac{\'e}p{\'e}}\ \emph {et~al.}(2020)\citenamefont
  {Sac{\'e}p{\'e}}, \citenamefont {Feigel’man},\ and\ \citenamefont
  {Klapwijk}}]{sacepe2020quantum}%
  \BibitemOpen
  \bibfield  {author} {\bibinfo {author} {\bibfnamefont {B.}~\bibnamefont
  {Sac{\'e}p{\'e}}}, \bibinfo {author} {\bibfnamefont {M.}~\bibnamefont
  {Feigel’man}},\ and\ \bibinfo {author} {\bibfnamefont {T.~M.}\ \bibnamefont
  {Klapwijk}},\ }\href@noop {} {\bibfield  {journal} {\bibinfo  {journal}
  {Nature Physics}\ }\textbf {\bibinfo {volume} {16}},\ \bibinfo {pages} {734}
  (\bibinfo {year} {2020})}\BibitemShut {NoStop}%
\bibitem [{\citenamefont {Calandra}\ \emph {et~al.}(2009)\citenamefont
  {Calandra}, \citenamefont {Mazin},\ and\ \citenamefont
  {Mauri}}]{calandra2009effect}%
  \BibitemOpen
  \bibfield  {author} {\bibinfo {author} {\bibfnamefont {M.}~\bibnamefont
  {Calandra}}, \bibinfo {author} {\bibfnamefont {I.}~\bibnamefont {Mazin}},\
  and\ \bibinfo {author} {\bibfnamefont {F.}~\bibnamefont {Mauri}},\
  }\href@noop {} {\bibfield  {journal} {\bibinfo  {journal} {Phys. Rev. B}\
  }\textbf {\bibinfo {volume} {80}},\ \bibinfo {pages} {241108} (\bibinfo
  {year} {2009})}\BibitemShut {NoStop}%
\bibitem [{\citenamefont {Lian}\ \emph {et~al.}(2018)\citenamefont {Lian},
  \citenamefont {Si},\ and\ \citenamefont {Duan}}]{lian2018unveiling}%
  \BibitemOpen
  \bibfield  {author} {\bibinfo {author} {\bibfnamefont {C.-S.}\ \bibnamefont
  {Lian}}, \bibinfo {author} {\bibfnamefont {C.}~\bibnamefont {Si}},\ and\
  \bibinfo {author} {\bibfnamefont {W.}~\bibnamefont {Duan}},\ }\href@noop {}
  {\bibfield  {journal} {\bibinfo  {journal} {Nano Lett.}\ }\textbf {\bibinfo
  {volume} {18}},\ \bibinfo {pages} {2924} (\bibinfo {year}
  {2018})}\BibitemShut {NoStop}%
\bibitem [{\citenamefont {Ugeda}\ \emph {et~al.}(2016)\citenamefont {Ugeda},
  \citenamefont {Bradley}, \citenamefont {Zhang}, \citenamefont {Onishi},
  \citenamefont {Chen}, \citenamefont {Ruan}, \citenamefont
  {Ojeda-Aristizabal}, \citenamefont {Ryu}, \citenamefont {Edmonds},
  \citenamefont {Tsai}, \citenamefont {Riss}, \citenamefont {{M}o},
  \citenamefont {Lee}, \citenamefont {Zettl}, \citenamefont {Hussain},
  \citenamefont {Shen},\ and\ \citenamefont
  {Crommie}}]{ugeda2016characterization}%
  \BibitemOpen
  \bibfield  {author} {\bibinfo {author} {\bibfnamefont {M.~M.}\ \bibnamefont
  {Ugeda}}, \bibinfo {author} {\bibfnamefont {A.~J.}\ \bibnamefont {Bradley}},
  \bibinfo {author} {\bibfnamefont {Y.}~\bibnamefont {Zhang}}, \bibinfo
  {author} {\bibfnamefont {S.}~\bibnamefont {Onishi}}, \bibinfo {author}
  {\bibfnamefont {Y.}~\bibnamefont {Chen}}, \bibinfo {author} {\bibfnamefont
  {W.}~\bibnamefont {Ruan}}, \bibinfo {author} {\bibfnamefont {C.}~\bibnamefont
  {Ojeda-Aristizabal}}, \bibinfo {author} {\bibfnamefont {H.}~\bibnamefont
  {Ryu}}, \bibinfo {author} {\bibfnamefont {M.~T.}\ \bibnamefont {Edmonds}},
  \bibinfo {author} {\bibfnamefont {H.-Z.}\ \bibnamefont {Tsai}}, \bibinfo
  {author} {\bibfnamefont {A.}~\bibnamefont {Riss}}, \bibinfo {author}
  {\bibfnamefont {S.-K.}\ \bibnamefont {{M}o}}, \bibinfo {author}
  {\bibfnamefont {D.}~\bibnamefont {Lee}}, \bibinfo {author} {\bibfnamefont
  {A.}~\bibnamefont {Zettl}}, \bibinfo {author} {\bibfnamefont
  {Z.}~\bibnamefont {Hussain}}, \bibinfo {author} {\bibfnamefont {Z.-X.}\
  \bibnamefont {Shen}},\ and\ \bibinfo {author} {\bibfnamefont
  {M.}~\bibnamefont {Crommie}},\ }\href@noop {} {\bibfield  {journal} {\bibinfo
   {journal} {Nat. Phys.}\ }\textbf {\bibinfo {volume} {12}},\ \bibinfo {pages}
  {92} (\bibinfo {year} {2016})}\BibitemShut {NoStop}%
\bibitem [{\citenamefont {Bianco}\ \emph {et~al.}(2020)\citenamefont {Bianco},
  \citenamefont {{M}onacelli}, \citenamefont {Calandra}, \citenamefont
  {Mauri},\ and\ \citenamefont {Errea}}]{bianco2020weak}%
  \BibitemOpen
  \bibfield  {author} {\bibinfo {author} {\bibfnamefont {R.}~\bibnamefont
  {Bianco}}, \bibinfo {author} {\bibfnamefont {L.}~\bibnamefont {{M}onacelli}},
  \bibinfo {author} {\bibfnamefont {M.}~\bibnamefont {Calandra}}, \bibinfo
  {author} {\bibfnamefont {F.}~\bibnamefont {Mauri}},\ and\ \bibinfo {author}
  {\bibfnamefont {I.}~\bibnamefont {Errea}},\ }\href@noop {} {\bibfield
  {journal} {\bibinfo  {journal} {Phys. Rev. Lett.}\ }\textbf {\bibinfo
  {volume} {125}},\ \bibinfo {pages} {106101} (\bibinfo {year}
  {2020})}\BibitemShut {NoStop}%
\bibitem [{\citenamefont {Leroux}\ \emph {et~al.}(2015)\citenamefont {Leroux},
  \citenamefont {Errea}, \citenamefont {Le~Tacon}, \citenamefont {Souliou},
  \citenamefont {Garbarino}, \citenamefont {Cario}, \citenamefont {Bosak},
  \citenamefont {Mauri}, \citenamefont {Calandra},\ and\ \citenamefont
  {Rodi{\`e}re}}]{leroux2015strong}%
  \BibitemOpen
  \bibfield  {author} {\bibinfo {author} {\bibfnamefont {M.}~\bibnamefont
  {Leroux}}, \bibinfo {author} {\bibfnamefont {I.}~\bibnamefont {Errea}},
  \bibinfo {author} {\bibfnamefont {M.}~\bibnamefont {Le~Tacon}}, \bibinfo
  {author} {\bibfnamefont {S.-M.}\ \bibnamefont {Souliou}}, \bibinfo {author}
  {\bibfnamefont {G.}~\bibnamefont {Garbarino}}, \bibinfo {author}
  {\bibfnamefont {L.}~\bibnamefont {Cario}}, \bibinfo {author} {\bibfnamefont
  {A.}~\bibnamefont {Bosak}}, \bibinfo {author} {\bibfnamefont
  {F.}~\bibnamefont {Mauri}}, \bibinfo {author} {\bibfnamefont
  {M.}~\bibnamefont {Calandra}},\ and\ \bibinfo {author} {\bibfnamefont
  {P.}~\bibnamefont {Rodi{\`e}re}},\ }\href@noop {} {\bibfield  {journal}
  {\bibinfo  {journal} {Phys. Rev. B}\ }\textbf {\bibinfo {volume} {92}},\
  \bibinfo {pages} {140303} (\bibinfo {year} {2015})}\BibitemShut {NoStop}%
\bibitem [{\citenamefont {Cho}\ \emph {et~al.}(2018)\citenamefont {Cho},
  \citenamefont {Ko{\'n}czykowski}, \citenamefont {Teknowijoyo}, \citenamefont
  {Tanatar}, \citenamefont {Guss}, \citenamefont {Gartin}, \citenamefont
  {Wilde}, \citenamefont {Kreyssig}, \citenamefont {McQueeney}, \citenamefont
  {Goldman} \emph {et~al.}}]{cho2018using}%
  \BibitemOpen
  \bibfield  {author} {\bibinfo {author} {\bibfnamefont {K.}~\bibnamefont
  {Cho}}, \bibinfo {author} {\bibfnamefont {M.}~\bibnamefont
  {Ko{\'n}czykowski}}, \bibinfo {author} {\bibfnamefont {S.}~\bibnamefont
  {Teknowijoyo}}, \bibinfo {author} {\bibfnamefont {M.~A.}\ \bibnamefont
  {Tanatar}}, \bibinfo {author} {\bibfnamefont {J.}~\bibnamefont {Guss}},
  \bibinfo {author} {\bibfnamefont {P.}~\bibnamefont {Gartin}}, \bibinfo
  {author} {\bibfnamefont {J.~M.}\ \bibnamefont {Wilde}}, \bibinfo {author}
  {\bibfnamefont {A.}~\bibnamefont {Kreyssig}}, \bibinfo {author}
  {\bibfnamefont {R.}~\bibnamefont {McQueeney}}, \bibinfo {author}
  {\bibfnamefont {A.~I.}\ \bibnamefont {Goldman}}, \emph {et~al.},\ }\href@noop
  {} {\bibfield  {journal} {\bibinfo  {journal} {Nat. Comm.}\ }\textbf
  {\bibinfo {volume} {9}},\ \bibinfo {pages} {1} (\bibinfo {year}
  {2018})}\BibitemShut {NoStop}%
\bibitem [{\citenamefont {Wang}\ \emph {et~al.}(2017)\citenamefont {Wang},
  \citenamefont {Huang}, \citenamefont {Lin}, \citenamefont {Cui},
  \citenamefont {Chen}, \citenamefont {Zhu}, \citenamefont {Liu}, \citenamefont
  {Zeng}, \citenamefont {Zhou}, \citenamefont {Yu} \emph
  {et~al.}}]{wang2017high}%
  \BibitemOpen
  \bibfield  {author} {\bibinfo {author} {\bibfnamefont {H.}~\bibnamefont
  {Wang}}, \bibinfo {author} {\bibfnamefont {X.}~\bibnamefont {Huang}},
  \bibinfo {author} {\bibfnamefont {J.}~\bibnamefont {Lin}}, \bibinfo {author}
  {\bibfnamefont {J.}~\bibnamefont {Cui}}, \bibinfo {author} {\bibfnamefont
  {Y.}~\bibnamefont {Chen}}, \bibinfo {author} {\bibfnamefont {C.}~\bibnamefont
  {Zhu}}, \bibinfo {author} {\bibfnamefont {F.}~\bibnamefont {Liu}}, \bibinfo
  {author} {\bibfnamefont {Q.}~\bibnamefont {Zeng}}, \bibinfo {author}
  {\bibfnamefont {J.}~\bibnamefont {Zhou}}, \bibinfo {author} {\bibfnamefont
  {P.}~\bibnamefont {Yu}}, \emph {et~al.},\ }\href@noop {} {\bibfield
  {journal} {\bibinfo  {journal} {Nat. Comm.}\ }\textbf {\bibinfo {volume}
  {8}},\ \bibinfo {pages} {1} (\bibinfo {year} {2017})}\BibitemShut {NoStop}%
\bibitem [{\citenamefont {Wickramaratne}\ \emph {et~al.}(2021)\citenamefont
  {Wickramaratne}, \citenamefont {Haim}, \citenamefont {Khodas},\ and\
  \citenamefont {Mazin}}]{nbse2_proximity}%
  \BibitemOpen
  \bibfield  {author} {\bibinfo {author} {\bibfnamefont {D.}~\bibnamefont
  {Wickramaratne}}, \bibinfo {author} {\bibfnamefont {M.}~\bibnamefont {Haim}},
  \bibinfo {author} {\bibfnamefont {M.}~\bibnamefont {Khodas}},\ and\ \bibinfo
  {author} {\bibfnamefont {I.~I.}\ \bibnamefont {Mazin}},\ }\href
  {https://doi.org/10.1103/PhysRevB.104.L060501} {\bibfield  {journal}
  {\bibinfo  {journal} {Phys. Rev. B}\ }\textbf {\bibinfo {volume} {104}},\
  \bibinfo {pages} {L060501} (\bibinfo {year} {2021})}\BibitemShut {NoStop}%
\bibitem [{\citenamefont {Bekaert}\ \emph {et~al.}(2020)\citenamefont
  {Bekaert}, \citenamefont {Khestanova}, \citenamefont {Hopkinson},
  \citenamefont {Birkbeck}, \citenamefont {Clark}, \citenamefont {Zhu},
  \citenamefont {Bandurin}, \citenamefont {Gorbachev}, \citenamefont
  {Fairclough}, \citenamefont {Zou} \emph {et~al.}}]{bekaert2020enhanced}%
  \BibitemOpen
  \bibfield  {author} {\bibinfo {author} {\bibfnamefont {J.}~\bibnamefont
  {Bekaert}}, \bibinfo {author} {\bibfnamefont {E.}~\bibnamefont {Khestanova}},
  \bibinfo {author} {\bibfnamefont {D.~G.}\ \bibnamefont {Hopkinson}}, \bibinfo
  {author} {\bibfnamefont {J.}~\bibnamefont {Birkbeck}}, \bibinfo {author}
  {\bibfnamefont {N.}~\bibnamefont {Clark}}, \bibinfo {author} {\bibfnamefont
  {M.}~\bibnamefont {Zhu}}, \bibinfo {author} {\bibfnamefont {D.~A.}\
  \bibnamefont {Bandurin}}, \bibinfo {author} {\bibfnamefont {R.}~\bibnamefont
  {Gorbachev}}, \bibinfo {author} {\bibfnamefont {S.}~\bibnamefont
  {Fairclough}}, \bibinfo {author} {\bibfnamefont {Y.}~\bibnamefont {Zou}},
  \emph {et~al.},\ }\href@noop {} {\bibfield  {journal} {\bibinfo  {journal}
  {Nano letters}\ }\textbf {\bibinfo {volume} {20}},\ \bibinfo {pages} {3808}
  (\bibinfo {year} {2020})}\BibitemShut {NoStop}%
\bibitem [{\citenamefont {Divilov}\ \emph {et~al.}(2021)\citenamefont
  {Divilov}, \citenamefont {Wan}, \citenamefont {Dreher}, \citenamefont
  {B{\"o}len}, \citenamefont {S{\'a}nchez-Portal}, \citenamefont {Ugeda},\ and\
  \citenamefont {Yndur{\'a}in}}]{divilov2021magnetic}%
  \BibitemOpen
  \bibfield  {author} {\bibinfo {author} {\bibfnamefont {S.}~\bibnamefont
  {Divilov}}, \bibinfo {author} {\bibfnamefont {W.}~\bibnamefont {Wan}},
  \bibinfo {author} {\bibfnamefont {P.}~\bibnamefont {Dreher}}, \bibinfo
  {author} {\bibfnamefont {E.}~\bibnamefont {B{\"o}len}}, \bibinfo {author}
  {\bibfnamefont {D.}~\bibnamefont {S{\'a}nchez-Portal}}, \bibinfo {author}
  {\bibfnamefont {M.~M.}\ \bibnamefont {Ugeda}},\ and\ \bibinfo {author}
  {\bibfnamefont {F.}~\bibnamefont {Yndur{\'a}in}},\ }\href@noop {} {\bibfield
  {journal} {\bibinfo  {journal} {Journal of Physics: Condensed Matter}\
  }\textbf {\bibinfo {volume} {33}},\ \bibinfo {pages} {295804} (\bibinfo
  {year} {2021})}\BibitemShut {NoStop}%
\bibitem [{\citenamefont {Das}\ and\ \citenamefont
  {Mazin}(2021)}]{das2021renormalized}%
  \BibitemOpen
  \bibfield  {author} {\bibinfo {author} {\bibfnamefont {S.}~\bibnamefont
  {Das}}\ and\ \bibinfo {author} {\bibfnamefont {I.~I.}\ \bibnamefont
  {Mazin}},\ }\href@noop {} {\bibfield  {journal} {\bibinfo  {journal} {arXiv
  preprint arXiv:2104.13205}\ } (\bibinfo {year} {2021})}\BibitemShut {NoStop}%
\bibitem [{\citenamefont {Wan}\ \emph {et~al.}(2021)\citenamefont {Wan},
  \citenamefont {Dreher}, \citenamefont {Harsh}, \citenamefont {Guinea},\ and\
  \citenamefont {Ugeda}}]{wan2021unconventional}%
  \BibitemOpen
  \bibfield  {author} {\bibinfo {author} {\bibfnamefont {W.}~\bibnamefont
  {Wan}}, \bibinfo {author} {\bibfnamefont {P.}~\bibnamefont {Dreher}},
  \bibinfo {author} {\bibfnamefont {R.}~\bibnamefont {Harsh}}, \bibinfo
  {author} {\bibfnamefont {F.}~\bibnamefont {Guinea}},\ and\ \bibinfo {author}
  {\bibfnamefont {M.~M.}\ \bibnamefont {Ugeda}},\ }\href@noop {} {\bibfield
  {journal} {\bibinfo  {journal} {arXiv preprint arXiv:2101.04050}\ } (\bibinfo
  {year} {2021})}\BibitemShut {NoStop}%
\bibitem [{\citenamefont {Dolgov}\ \emph {et~al.}(2005)\citenamefont {Dolgov},
  \citenamefont {Mazin}, \citenamefont {Golubov}, \citenamefont {Savrasov},\
  and\ \citenamefont {Maksimov}}]{MME}%
  \BibitemOpen
  \bibfield  {author} {\bibinfo {author} {\bibfnamefont {O.~V.}\ \bibnamefont
  {Dolgov}}, \bibinfo {author} {\bibfnamefont {I.~I.}\ \bibnamefont {Mazin}},
  \bibinfo {author} {\bibfnamefont {A.~A.}\ \bibnamefont {Golubov}}, \bibinfo
  {author} {\bibfnamefont {S.~Y.}\ \bibnamefont {Savrasov}},\ and\ \bibinfo
  {author} {\bibfnamefont {E.~G.}\ \bibnamefont {Maksimov}},\ }\href
  {https://doi.org/10.1103/PhysRevLett.95.257003} {\bibfield  {journal}
  {\bibinfo  {journal} {Phys. Rev. Lett.}\ }\textbf {\bibinfo {volume} {95}},\
  \bibinfo {pages} {257003} (\bibinfo {year} {2005})}\BibitemShut {NoStop}%
\bibitem [{\citenamefont {Heil}\ \emph {et~al.}(2017)\citenamefont {Heil},
  \citenamefont {Ponc{\'e}}, \citenamefont {Lambert}, \citenamefont {Schlipf},
  \citenamefont {Margine},\ and\ \citenamefont {Giustino}}]{heil2017origin}%
  \BibitemOpen
  \bibfield  {author} {\bibinfo {author} {\bibfnamefont {C.}~\bibnamefont
  {Heil}}, \bibinfo {author} {\bibfnamefont {S.}~\bibnamefont {Ponc{\'e}}},
  \bibinfo {author} {\bibfnamefont {H.}~\bibnamefont {Lambert}}, \bibinfo
  {author} {\bibfnamefont {M.}~\bibnamefont {Schlipf}}, \bibinfo {author}
  {\bibfnamefont {E.~R.}\ \bibnamefont {Margine}},\ and\ \bibinfo {author}
  {\bibfnamefont {F.}~\bibnamefont {Giustino}},\ }\href@noop {} {\bibfield
  {journal} {\bibinfo  {journal} {Phys. Rev. Lett}\ }\textbf {\bibinfo {volume}
  {119}},\ \bibinfo {pages} {087003} (\bibinfo {year} {2017})}\BibitemShut
  {NoStop}%
\bibitem [{\citenamefont {Margine}\ and\ \citenamefont {et. al.}()}]{margine}%
  \BibitemOpen
  \bibfield  {author} {\bibinfo {author} {\bibfnamefont {E.}~\bibnamefont
  {Margine}}\ and\ \bibinfo {author} {\bibnamefont {et. al.}},\ }\href@noop {}
  {}\bibinfo {howpublished} {"In preparation"}\BibitemShut {NoStop}%
\bibitem [{\citenamefont {Anikin}\ \emph {et~al.}(2020)\citenamefont {Anikin},
  \citenamefont {Schaller}, \citenamefont {Wiederrecht}, \citenamefont
  {Margine}, \citenamefont {Mazin},\ and\ \citenamefont {Karapetrov}}]{Goran}%
  \BibitemOpen
  \bibfield  {author} {\bibinfo {author} {\bibfnamefont {A.}~\bibnamefont
  {Anikin}}, \bibinfo {author} {\bibfnamefont {R.~D.}\ \bibnamefont
  {Schaller}}, \bibinfo {author} {\bibfnamefont {G.~P.}\ \bibnamefont
  {Wiederrecht}}, \bibinfo {author} {\bibfnamefont {E.~R.}\ \bibnamefont
  {Margine}}, \bibinfo {author} {\bibfnamefont {I.~I.}\ \bibnamefont {Mazin}},\
  and\ \bibinfo {author} {\bibfnamefont {G.}~\bibnamefont {Karapetrov}},\
  }\href {https://doi.org/10.1103/PhysRevB.102.205139} {\bibfield  {journal}
  {\bibinfo  {journal} {Phys. Rev. B}\ }\textbf {\bibinfo {volume} {102}},\
  \bibinfo {pages} {205139} (\bibinfo {year} {2020})}\BibitemShut {NoStop}%
\bibitem [{\citenamefont {Peng}\ \emph {et~al.}(2018)\citenamefont {Peng},
  \citenamefont {Yu}, \citenamefont {Wu}, \citenamefont {Zhou}, \citenamefont
  {Guo}, \citenamefont {Li}, \citenamefont {Zhao}, \citenamefont {Wu},\ and\
  \citenamefont {Xie}}]{peng2018disorder}%
  \BibitemOpen
  \bibfield  {author} {\bibinfo {author} {\bibfnamefont {J.}~\bibnamefont
  {Peng}}, \bibinfo {author} {\bibfnamefont {Z.}~\bibnamefont {Yu}}, \bibinfo
  {author} {\bibfnamefont {J.}~\bibnamefont {Wu}}, \bibinfo {author}
  {\bibfnamefont {Y.}~\bibnamefont {Zhou}}, \bibinfo {author} {\bibfnamefont
  {Y.}~\bibnamefont {Guo}}, \bibinfo {author} {\bibfnamefont {Z.}~\bibnamefont
  {Li}}, \bibinfo {author} {\bibfnamefont {J.}~\bibnamefont {Zhao}}, \bibinfo
  {author} {\bibfnamefont {C.}~\bibnamefont {Wu}},\ and\ \bibinfo {author}
  {\bibfnamefont {Y.}~\bibnamefont {Xie}},\ }\href@noop {} {\bibfield
  {journal} {\bibinfo  {journal} {ACS nano}\ }\textbf {\bibinfo {volume}
  {12}},\ \bibinfo {pages} {9461} (\bibinfo {year} {2018})}\BibitemShut
  {NoStop}%
\bibitem [{\citenamefont {Li}\ \emph {et~al.}(2017)\citenamefont {Li},
  \citenamefont {Deng}, \citenamefont {Wang}, \citenamefont {Liu},
  \citenamefont {Abeykoon}, \citenamefont {Dooryhee}, \citenamefont {Tomic},
  \citenamefont {Huang}, \citenamefont {Warren}, \citenamefont {Bozin} \emph
  {et~al.}}]{li2017superconducting}%
  \BibitemOpen
  \bibfield  {author} {\bibinfo {author} {\bibfnamefont {L.}~\bibnamefont
  {Li}}, \bibinfo {author} {\bibfnamefont {X.}~\bibnamefont {Deng}}, \bibinfo
  {author} {\bibfnamefont {Z.}~\bibnamefont {Wang}}, \bibinfo {author}
  {\bibfnamefont {Y.}~\bibnamefont {Liu}}, \bibinfo {author} {\bibfnamefont
  {M.}~\bibnamefont {Abeykoon}}, \bibinfo {author} {\bibfnamefont
  {E.}~\bibnamefont {Dooryhee}}, \bibinfo {author} {\bibfnamefont
  {A.}~\bibnamefont {Tomic}}, \bibinfo {author} {\bibfnamefont
  {Y.}~\bibnamefont {Huang}}, \bibinfo {author} {\bibfnamefont {J.~B.}\
  \bibnamefont {Warren}}, \bibinfo {author} {\bibfnamefont {E.~S.}\
  \bibnamefont {Bozin}}, \emph {et~al.},\ }\href@noop {} {\bibfield  {journal}
  {\bibinfo  {journal} {npj Quantum Materials}\ }\textbf {\bibinfo {volume}
  {2}},\ \bibinfo {pages} {1} (\bibinfo {year} {2017})}\BibitemShut {NoStop}%
\bibitem [{\citenamefont {Luo}\ \emph {et~al.}(2015)\citenamefont {Luo},
  \citenamefont {Xie}, \citenamefont {Tao}, \citenamefont {Inoue},
  \citenamefont {Gyenis}, \citenamefont {Krizan}, \citenamefont {Yazdani},
  \citenamefont {Zhu},\ and\ \citenamefont {Cava}}]{luo2015polytypism}%
  \BibitemOpen
  \bibfield  {author} {\bibinfo {author} {\bibfnamefont {H.}~\bibnamefont
  {Luo}}, \bibinfo {author} {\bibfnamefont {W.}~\bibnamefont {Xie}}, \bibinfo
  {author} {\bibfnamefont {J.}~\bibnamefont {Tao}}, \bibinfo {author}
  {\bibfnamefont {H.}~\bibnamefont {Inoue}}, \bibinfo {author} {\bibfnamefont
  {A.}~\bibnamefont {Gyenis}}, \bibinfo {author} {\bibfnamefont {J.~W.}\
  \bibnamefont {Krizan}}, \bibinfo {author} {\bibfnamefont {A.}~\bibnamefont
  {Yazdani}}, \bibinfo {author} {\bibfnamefont {Y.}~\bibnamefont {Zhu}},\ and\
  \bibinfo {author} {\bibfnamefont {R.~J.}\ \bibnamefont {Cava}},\ }\href@noop
  {} {\bibfield  {journal} {\bibinfo  {journal} {Proceedings of the National
  Academy of Sciences}\ }\textbf {\bibinfo {volume} {112}},\ \bibinfo {pages}
  {E1174} (\bibinfo {year} {2015})}\BibitemShut {NoStop}%
\bibitem [{\citenamefont {Bl{\"o}chl}(1994)}]{Blochl_PAW}%
  \BibitemOpen
  \bibfield  {author} {\bibinfo {author} {\bibfnamefont {P.~E.}\ \bibnamefont
  {Bl{\"o}chl}},\ }\href@noop {} {\bibfield  {journal} {\bibinfo  {journal}
  {Phys. Rev. B}\ }\textbf {\bibinfo {volume} {50}},\ \bibinfo {pages} {17953}
  (\bibinfo {year} {1994})}\BibitemShut {NoStop}%
\bibitem [{\citenamefont {Kresse}\ and\ \citenamefont
  {Hafner}(1993)}]{VASP_ref}%
  \BibitemOpen
  \bibfield  {author} {\bibinfo {author} {\bibfnamefont {G.}~\bibnamefont
  {Kresse}}\ and\ \bibinfo {author} {\bibfnamefont {J.}~\bibnamefont
  {Hafner}},\ }\href@noop {} {\bibfield  {journal} {\bibinfo  {journal} {Phys.
  Rev. B}\ }\textbf {\bibinfo {volume} {47}},\ \bibinfo {pages} {558} (\bibinfo
  {year} {1993})}\BibitemShut {NoStop}%
\bibitem [{\citenamefont {Kresse}\ and\ \citenamefont
  {Furthm{\"u}ller}(1996)}]{VASP_ref2}%
  \BibitemOpen
  \bibfield  {author} {\bibinfo {author} {\bibfnamefont {G.}~\bibnamefont
  {Kresse}}\ and\ \bibinfo {author} {\bibfnamefont {J.}~\bibnamefont
  {Furthm{\"u}ller}},\ }\href@noop {} {\bibfield  {journal} {\bibinfo
  {journal} {Phys. Rev. B}\ }\textbf {\bibinfo {volume} {54}},\ \bibinfo
  {pages} {11169} (\bibinfo {year} {1996})}\BibitemShut {NoStop}%
\bibitem [{\citenamefont {Perdew}\ \emph {et~al.}(1996)\citenamefont {Perdew},
  \citenamefont {Burke},\ and\ \citenamefont
  {Ernzerhof}}]{perdew1996generalized}%
  \BibitemOpen
  \bibfield  {author} {\bibinfo {author} {\bibfnamefont {J.~P.}\ \bibnamefont
  {Perdew}}, \bibinfo {author} {\bibfnamefont {K.}~\bibnamefont {Burke}},\ and\
  \bibinfo {author} {\bibfnamefont {M.}~\bibnamefont {Ernzerhof}},\ }\href@noop
  {} {\bibfield  {journal} {\bibinfo  {journal} {Phys. Rev. Lett.}\ }\textbf
  {\bibinfo {volume} {77}},\ \bibinfo {pages} {3865} (\bibinfo {year}
  {1996})}\BibitemShut {NoStop}%
\bibitem [{\citenamefont {Sandratskii}(1991)}]{sandratskii1991symmetry}%
  \BibitemOpen
  \bibfield  {author} {\bibinfo {author} {\bibfnamefont {L.}~\bibnamefont
  {Sandratskii}},\ }\href@noop {} {\bibfield  {journal} {\bibinfo  {journal}
  {Journal of Physics: Condensed Matter}\ }\textbf {\bibinfo {volume} {3}},\
  \bibinfo {pages} {8565} (\bibinfo {year} {1991})}\BibitemShut {NoStop}%
\end{thebibliography}


\end{document}